\documentstyle[12pt]{article}
\begin {document}
\begin{titlepage}
October 1992 \hfill HU Berlin--IEP--92/6 \\
\mbox{}\hfill hep-th/9210159
\vspace{6ex}
\Large
\begin {center}
\bf{Non critical super strings on world sheets of constant curvature}
\end {center}
\large
\vspace{3ex}
\begin{center}
Stefan F\"orste\footnote{email: foerste@ifh.de}
\end{center}
\normalsize
\it
\vspace{3ex}
\begin{center}
Fachbereich Physik der Humboldt--Universit\"at \\
Institut f\"ur Elementarteilchenphysik \\
Invalidenstra\ss e 110, O--1040 Berlin
\end{center}
\vspace{6 ex }
\rm
\begin{center}
\bf{Abstract}
\end{center}
We consider correlation functions in Neveu--Schwarz string theory coupled
to two dimensional gravity. The action for the 2D gravity consists of the
string induced Liouville action and the Jackiw--Teitelboim action
describing pure 2D gravity. Then gravitational dressed dimensions of vertex
operators are equal to their bare conformal dimensions. There are two
possible interpretations of the model. Considering the 2D dilaton and the
Liouville field as additional target space coordinates one gets a
$d+2$-dimensional critical string. In the $d$-dimensional non critical
string picture gravitational fields retain their original meaning and
for $d=4$ one can get a mass spectrum via consistency requirements.
In both cases a GSO projection is possible.
\end {titlepage}

\section{Introduction}

One serious problem of super string theory is that observables depend on the
vielbein describing the world sheet swept out by the string unless the
dimension $d$ of the target space is ten. Polyakov's proposal \cite{poly}
to include two dimensional quantum gravity, i.e.\ to integrate over all
vielbeine, does not solve the problem completely because in target space
dimensions between one and ten the partition function has a complex scaling
dimension, (complex string susceptibility), \cite{kpz,apiky}.
A slight modification of that approach consists in the inclusion of a
classical action for two dimensional super gravity.
The Einstein--Hilbert action is not a suitable candidate since it is a
topological constant in two dimensions.
We will use the super symmetric version of the Jackiw--Teitelboim action
\cite{jacky,chamwy}\footnote{Possible  generalizations of the
Jackiw--Teitelboim action are considered in \cite{lio}.}
\begin{equation}     \label{sjt}
S_{JT}=-\frac{1}{\pi }\int d^{2}Z\, E\Phi \left( R_{+-}+H\right),
\end{equation}
where $H$ is a cosmological constant, $R_{+-}$ is the super curvature,
$\Phi$ a super scalar field which can be considered as a partner of the
vielbein field and is sometimes called a 2D dilaton \cite{chamsnu},
and $E$ is the super determinant of the inverse vielbein
($E_{A}=E_{A}^{M}\partial_{M}$)
\begin{equation}
E=s\det E^{A}_{M}.
\end{equation}
Throughout the present paper we use conventions of \cite{dhoker}. The super
scalar $\Phi$ enters the action like a Lagrangean multiplier and thus
(\ref{sjt}) represents the constraint of constant curvature
\begin{equation}                                           \label{constr}
R_{+-}+H=0.
\end{equation}
Super string theory coupled to the 2D gravity (\ref{sjt}) was considered in
\cite{chamsnu,chamslet,odi}. In \cite{chamslet} it was shown that the inclusion
of the
Jackiw--Teitelboim action provides a real string susceptibility for every
target space dimension $d$.
As a next step it is reasonable to consider
correlation functions. The second section of our paper is addressed to the
calculation of the N--point tachyon amplitude. One can read of the
gravitational dressed dimensions of vertex operators from the $H$--dependence
of the N--point function. In the third section we will treat $\Phi$,
and the Liouville field $\Sigma$ as additional target space coordinates.
Analyzing the pole structure of the integrated N--point function we get the
mass spectrum of the theory. The fields $\Phi$ and $\Sigma$ retain their
original meaning in the last section. For $d=4$ it is possible to find
a mass spectrum via consistency requirements.

\section{Non critical amplitudes}
\setcounter{equation}{0}
Before considering the N-point function we describe briefly the geometry of
the world sheet. More details are given in \cite{dhoker,west,howe}.
The world sheet is a two dimensional super manifold, i.e.\ a two
dimensional surface where on each point a two dimensional space
consisting of  Grassmann numbers is attached as a fiber. In order
to avoid redundant  degrees of freedom one puts certain constraints on the
torsion. Using the Bianchi identities their remains only one
field describing the curvature \cite{howe} which is conveniently chosen
to be one component of the Ricci tensor,
\begin{equation}
R_{+-}.
\end{equation}
In our paper we will confine ourself to simply connected world sheets,
(topological terms refer to the surface described by commuting
coordinates because the fibers possess the trivial topology).
Then it is possible to perform local Lorentz transformations, and
diffeomorphisms in such a way that the vielbein is given in a
super conformal flat form \cite{west,howe},
\begin{equation}                       \label{skfgauge}
E_{+}=e^{-\frac{\Sigma (Z)}{2}}D_{+}\equiv e^{-\frac{\Sigma}{2}}\left(
\partial_{\theta }+\theta \partial_{z}\right),
\end{equation}
\begin{equation}
E_{-}=e^{-\frac{\Sigma (Z)}{2}}D_{-}\equiv e^{-\frac{\Sigma}{2}}\left(
\partial_{\bar{\theta} }+\bar{\theta}\partial_{\bar{z}}\right).
\end{equation}
(Small Latins denote commuting coordinates, small Greeks anti commuting
coordinates. Both types are commonly characterized by capital Latins.)
The constraints on the torsion determine all the other vielbeine
\cite{dhoker,west},
\begin{equation}    \label{equad}
E_{z}=E_{+}E_{+},\; \; E_{\bar{z}}=E_{-}E_{-}.
\end{equation}
We will use the super field formalism,
\begin{equation}
F(Z)=F(z,\bar{z},\theta ,\bar{\theta })=f(z,\bar{z})+\theta
\varphi ^{(+)}(z,\bar{z})+\bar{\theta }\varphi ^{(-)}(z,\bar{z})
+\theta \bar{\theta }b(z,\bar{z}).
\end{equation}
In non critical dimensions a normal ordered vertex operator contains
a covariant cutoff and a super scalar density \cite{apiky,dudi,do}.
The tachyon vertex is for example given by (cf.\ also \cite{me})
\begin{equation}
T_{j}(Z_{j})=:e^{ik_{j}X(Z_{j})}:=e^{ik_{j}X(Z_{j})}B_{j}(Z_{j}),
\end{equation}
\begin{equation}
B_{j}(Z_{j})=\left( \epsilon e^{\Sigma (Z_{j})}\right)^{\beta_{j}-1}
e^{\Sigma (Z_{j})+i\gamma_{j}\Phi (Z_{j})}.
\end{equation}
Here $\epsilon$ is a UV cutoff and $\beta_{j}, \gamma_{j}$ are determined
by the requirement that the final result should be finite for
$\epsilon \longrightarrow 0$. It is reasonable to restrict to the special
case that the scalar density contains only the vielbein field,
i.e.\ $\gamma_{j}=0$. However, later on we will interpret $\Phi$, and
$\Sigma$ as additional target space coordinates. Then $B_{j}$ is just
a part of the vertex and $\gamma_{j}=0$ is no longer a reasonable special
case. Therefore we will set $\gamma_{j}=0$ only when we are considering
the non critical string.

Now we are interested in the N--point tachyon function,
\begin{equation}
\langle \prod_{j=1}^{N}T_{j}(Z_{j})\rangle =\frac{1}{\cal Z}
\int {\cal D}E{\cal D}\Phi {\cal D}Xe^{-S_{JT}-S_{M}}\prod_{j=1}^{N}
T_{j}(Z_{j}),
\end{equation}
where
\begin{equation}           \label{susy-action}
S_{M}=\frac{1}{4\pi }\int d^{2}Z E {\cal D}_{-}X^{\mu }(Z)
{\cal D}_{+}X_{\mu }(Z),
\; \; \mu =1,\ldots ,d
\end{equation}
is the super string action. The integral over all vielbein components
can be expressed as an integral over local Lorentz transformations,
diffeomorphisms, and over the Weyl factor $\Sigma$. There are no anomalies
in the Lorentz transformations, and the diffeomorphisms, hence those
integrals provide an uninteresting factor. Therefore we replace the
vielbein integral by an integration over the Weyl factor $\Sigma$ times
a Jacobian,
\begin{equation}
{\cal D}E ={\cal D}_{E}\Sigma \left(s\det PP^{\dagger }\right)^{\frac{1}{2}}.
\end{equation}
The $\Sigma$ dependence of the Jacobian was determined in \cite{mart},
\begin{equation}
\left( s\det PP^{\dagger}\right) ^{\frac{1}{2}}\sim
e^{-10S_{sL}},
\end{equation}
where $S_{sL}$ is the super Liouville action,
\begin{equation}
S_{sL}=\frac{1}{8\pi }\int d^{2}ZD_{-}\Sigma D_{+}\Sigma.
\end{equation}
(In the bosonic case one has to introduce a cosmological term in order
to get a renormalizable theory \cite{oalv}. Anyway, an inclusion of a
cosmological term would not change essential results (cf.\ \cite{me}).)
Performing the $X$ integrals we get
\begin{equation}                             \label{smatterpart}
\langle \prod_{j=1}^{N}T_{j}(Z_{j})\rangle
=\delta^{(d)}\left( \sum_{j=1}^{N}k_{j}\right)
\prod_{i\not= j}^{N}\left| Z_{ij}\right| ^{k_{i}k_{j}}\left(
\frac{\epsilon }{\mu }\right)^{\sum_{j=1}^{N}k_{j}^{2}}
\langle \langle e^{-(10-d)S_{sL}}\prod_{j=1}^{N}B_{j}(Z_{j})\rangle \rangle .
\end{equation}
We have regularized
$$ \log (0)\longrightarrow \log \left(\frac{\epsilon }{\mu }\right), $$
where $\epsilon$ is a UV cutoff and $\mu $ is a renormalization group
scale. Furthermore we have used the notation of super space distance,
\begin{equation}
Z_{ij}\equiv Z_{i}-Z_{j}=z_{i}-z_{j}-\theta_{i}\theta_{j}.
\end{equation}
The remaining gravitational expectation value is defined by
\begin{equation}
\langle \langle \cdots \rangle \rangle =
\frac{1}{\cal Z}\int {\cal D}_{E}\Sigma {\cal D}_{E}\Phi
\, e^{-S_{JT}}\cdots \; .
\end{equation}
The functional measures are given by the requirement that the Gaussian
integral is normalized to one \cite{dhoker,apiky} (cf.\ also \cite{dudi}),
\begin{equation}
\int {\cal D}_{E}\delta \Lambda e^{-(\delta \Lambda ,\delta \Lambda )_{E}}=1,
\end{equation}
\begin{equation}
(\delta \Lambda ,\delta \Lambda )_{E}=\int d^{2}Z\, E \left( \delta \Lambda
\right)^{2}=\int d^{2}Z\, e^{\Sigma (Z)}\left( \delta \Lambda \right)^{2}.
\end{equation}
Hence the $\Sigma$ measure is not translation invariant. Therefore we split
$\Sigma$ into a quantum part $\Sigma$, and into a classical background part
$\hat{\Sigma}$,
$$\Sigma \longrightarrow \Sigma +\hat{\Sigma}$$
and use measures referring to
\begin{equation}               \label{hutbein}
\hat{E_{\pm}}=e^{-\frac{\hat{\Sigma}}{2}}D_{\pm}.
\end{equation}
The calculation of the arising Jacobian is performed in \cite{mavro}.
Taking into account the full $\hat{\Sigma}$ dependence we get
\begin{eqnarray}                                 \label{sprick}
\lefteqn{\langle \langle \prod_{j=1}^{N}e^{\beta_{j}\Sigma (Z_{j})+i\gamma
_{j}\Phi (Z_{j})}\rangle \rangle =}\nonumber \\
& & \frac{1}{\cal Z}e^{-(10-d)S_{sL}[\hat{\Sigma}]}\prod_{j=1}^{N}
e^{\beta_{j}\hat{\Sigma}(Z_{j})}
\nonumber \\
& &
\int {\cal D}_{\hat{E}}\Sigma {\cal D}_{\hat{E}}\Phi \,
e^{-\hat{S}_{sL}-\hat{S}_{JT}}\prod_{j=1}^{N}
e^{\beta_{j}\Sigma (Z_{j})+i\gamma_{j}\Phi (Z_{j})},
\end{eqnarray}
where
\begin{equation}                                             \label{hat-sl}
\hat{S}_{sL}=\frac{a}{\pi }\int d^{2}Z\, \hat{E}\left(  \hat{\cal D}_{-}
\Sigma \hat{\cal D}_{+}\Sigma +i\hat{R}_{+-}\Sigma\right),
\end{equation}
\begin{equation}
a=\frac{8-d}{8},
\end{equation}
and
\begin{equation}                                               \label{hat-sjt}
\hat{S}_{JT}=\frac{i}{\pi }\int d^{2}Z\, \hat{E}\left( 2\hat{\cal D}_{-}\Phi
\hat{\cal D}_{+}\Sigma +i\hat{R}_{+-}\Phi +iH\left( \Phi e^{\Sigma}\right)
_{ren}\right).
\end{equation}
We note that in the case without Jackiw--Teitelboim  action one gets
$8a=9-d$\ because in that case a contribution from the $\Phi$ measure
is absent. Furthermore we have used \cite{chamslet}
\begin{equation}
R_{+-}=e^{-\Sigma}\left( \hat{R}_{+-}-2i\hat{\cal D}_{+}\hat{\cal D}_{-}
\Sigma \right),
\end{equation}
where the hat refers to the background vielbein $\hat{E}$ given in
(\ref{hutbein}). In (\ref{hat-sjt}) we have admitted a renormalization of
the exponential term. First let us consider the case $H=0$.
It is convenient to use
\begin{equation}                   \label{redef}
\Psi =\Sigma +\frac{i}{a}\Phi
\end{equation}
instead of $\Sigma$. Then the theory is described by the sum of two independent
terms,
\begin{eqnarray}                \label{superfrei}
\hat{S}_{sL}+\hat{S}_{JT|H=0} & = &
\frac{a}{\pi }\int d^{2}Z\, \hat{E}\left( \hat{\cal D}_{-}\Psi
\hat{\cal D}_{+}\Psi +i\hat{R}_{+-}\Psi \right)+
\nonumber \\
& & \frac{1}{a\pi }\int d^{2}Z\, \hat{E}\hat{\cal D}_{-}\Phi
\hat{\cal D}_{+}\Phi.
\end{eqnarray}
That is a super conformal field theory with central charge
\begin{equation}
c=8a+1+1=10-d.
\end{equation}
(A review about super conformal field theory is given in \cite{fried}.)
Together with the matter part (and the gauge fixing ghost part) we have
a vanishing total central charge, i.e.\ a conformal invariant theory not
depending on the arbitraryly chosen background field $\hat{\Sigma}$. In order
to ensure that conformal invariance is not spoiled for $H\not= 0$ we require
the renormalized exponential term to be a primary field of dimension
one half. The super conformal dimension of a general primary field
is given by
\begin{eqnarray}     \label{skondim}
\Delta \left( e^{\omega \Sigma +i\gamma \Phi}\right) & = &
\Delta \left( e^{\omega \Psi}\right)+\Delta \left( e^{i\left( \gamma -
\frac{\omega}{a}\right)\Phi}\right) \nonumber \\
& = & \frac{\omega}{2}-\frac{\omega^{2}}{8a}+\frac{a}{8}\left( \gamma -
\frac{\omega}{a}\right) ^{2} \nonumber \\
& = & \frac{\omega}{2}-\frac{\gamma \omega }{4}+\frac{a}{8}\gamma ^{2}.
\end{eqnarray}
Hence the condition $2\Delta =1$ does not provide a unique renormalization
prescription. Moreover there are also primaries not contained in the
exponential ansatz (\ref{skondim}) \cite{chamsnu}. As in the bosonic case
\cite{me} we require in addition to $2\Delta =1$ that renormalized and not
renormalized operators coincide in the semi classical limit
($d\longrightarrow -\infty$). (That is a slight modification of the
argumentation in the case without $S_{JT}$. There one uses the non
existence of the semi classical limit to exclude one of two possible
solutions \cite{dudi}.)
We get the following renormalization prescription,
\begin{equation}                                         \label{sren}
\left( \Phi e^{\Sigma}\right)_{ren}=-\frac{ia}{2}e^{\Sigma}
\left( e^{\frac{2i}{a}\Phi}-1\right)=\Phi e^{\Sigma}+o\left( \frac{1}{a}
\right).
\end{equation}
In terms of super fields the Gauss--Bonnet theorem is given by
\cite{chamslet,dhoker}
\begin{equation}   \label{supergau}
\frac{i}{4\pi }\int d^{2}Z\, E R_{+-}=1-h,
\end{equation}
where $h$ is the genus of  the world sheet.
Integrating out the the zero modes in (\ref{sprick}) and
neglecting uninteresting factors we get
\begin{eqnarray}                     \label{sprick1}
\lefteqn{\langle \langle \prod_{j=1}^{N}e^{\beta_{j}\Sigma (Z_{j})+
i\gamma_{j}\Phi (Z_{j})}\rangle \rangle =}\nonumber \\
& & \frac{1}{\cal Z}\Gamma \left(-t\right)\Gamma\left(-s\right)  H^{t+s}
e^{-(10-d)S[\hat{\Sigma}]}\prod_{j=1}^{N}e^{\beta_{j}\hat{\Sigma}
(Z_{j})} \nonumber \\
& & \int {\cal D}_{\hat{E}\perp}\Sigma {\cal D}_{\hat{E}\perp}\Phi
\, e^{-\hat{S}_{JT\mid H=0}-\hat{S}_{sL}} F^{t}A^{s}
\prod_{j=1}^{N}e^{\beta_{j}\Sigma (Z_{j})+i\gamma_{j}\Phi (Z_{j})},
\end{eqnarray}
where the $\perp $ label at measures indicates that zero mode
integration has been performed. We used the following abbreviations,
\begin{equation}
F=\int d^{2}Z\, \hat{E}e^{\Sigma +\frac{2i}{a}\Phi},
\end{equation}
\begin{equation}
A=\int d^{2}Z\, \hat{E}e^{\Sigma},
\end{equation}
\begin{equation}                 \label{st=}
t=2a-\frac{a}{2}\sum_{j=1}^{N}\gamma_{j},
\end{equation}
\begin{equation}                \label{ss=}
s=4a-t-\sum_{j=1}^{N}\beta_{j}=2a+\frac{a}{2}\sum_{j=1}^{N}\gamma_{j}
-\sum_{j=1}^{N}\beta_{j}.
\end{equation}
The scaling behavior of the partition function is
\begin{equation} \label{su-ska-zu}
{\cal Z}\sim H^{4a}=H^{\frac{1}{2}(8-d)}.
\end{equation}
This coincides with the result obtained by a calculation using non
translation invariant measures \cite{chamslet}.
A convenient covariant definition of zero modes is given by \cite{do}
\begin{equation}   \label{snulldef}
\Sigma_{0}=\frac{i}{4\pi}\int d^{2}Z\, \hat{E}\hat{R}_{+-}\Sigma ,
\end{equation}
\begin{equation}
\Phi_{0}=\frac{i}{4\pi}\int d^{2}Z\, \hat{E}\hat{R}_{+-}\Phi .
\end{equation}
Redefining the fields according to (\ref{redef}) we get for $\Psi$, and $\Phi$
the propagator
\begin{equation}
G(Z_{j},Z_{k}|\hat{\Sigma})=-\log \left| Z_{jk}\right| -\frac{1}{2}
\hat{\Sigma} (Z_{j})-\frac{1}{2}\hat{\Sigma}(Z_{k})+2S_{sL}[\hat{\Sigma} ].
\end{equation}
A further calculation is possible for $s$, and $t$ being non negative
integers. In that case we obtain
\begin{eqnarray}       \label{sprick2}
\lefteqn{\langle \langle \prod_{j=1}^{N} e^{\beta_{j}\Sigma (Z_{j})+
i\gamma_{j}\Phi (Z_{j})}\rangle \rangle =}\nonumber \\
& & \frac{1}{\cal Z}H^{t+s}\Gamma (-t)\Gamma (-s)
e^{-(8-d)S_{sL}[\hat{\Sigma}]}\prod_{j=1}^{N}e^{\beta_{j}\hat{\Sigma}(Z_{j}}
\nonumber \\
& & \left( \int \prod_{j=N+1}^{N+t+s}d^{2}Z_{j}\,
e^{\beta_{j}\hat{\Sigma}(Z_{j})}\right) \prod_{j=1}^{N+t+s}
\prod_{k=1}^{N+t+s}e^{(\gamma_{j}\beta_{k}+\gamma_{k}\beta_{j}-a\gamma_{j}
\gamma_{k})\frac{1}{4}G(Z_{j},Z_{k}|\hat{\Sigma})},
\end{eqnarray}                     
where
\begin{equation}
\beta_{N+1}=\ldots =\beta_{N+t+s}=1,
\end{equation}
\begin{equation}
\gamma_{N+1}=\ldots =\gamma_{N+t}=\frac{2}{a},
\end{equation}
\begin{equation}
\gamma_{N+t+1}=\ldots =\gamma_{N+t+s}=0.
\end{equation}
With
\begin{eqnarray}   \label{sneutra}
\sum_{j=1}^{N+t+s}\gamma_{j}=\sum_{j=1}^{N}\gamma_{j}+\frac{2}{a}t
& = & 4, \nonumber \\
\sum_{j=1}^{N+t+s}\beta_{j}=\sum_{j=1}^{N}\beta_{j}+t+s
& = & 4a
\end{eqnarray}
it is easy to show that the $\hat{\Sigma}$ dependence drops out.
Finally we get for the N--point tachyon function
\begin{eqnarray}                  \label{superresult}
\lefteqn{\langle \prod_{j=1}^{N}T_{j}(Z_{j}) \rangle =}\nonumber \\
& & \delta^{(d)}\left( \sum_{j=1}^{N}k_{j}\right)
H^{t+s-4a}\frac{\Gamma (-t)\Gamma (-s)}{\Gamma (-2a)\Gamma (-2a)}
\left( \frac{\epsilon }{\mu }
\right)^{\sum_{j=1}^{N}\left( k_{j}^{2}+ \frac{a}{4}
\gamma_{j}^{2}-\frac{\gamma_{j}\beta_{j}}{2}\right)}
\epsilon^{\sum_{j=1}^{N}(\beta_{j}-1)}
\nonumber \\
& & \int \prod_{A=1}^{t}d^{2}W_{A}\, \prod_{j=1}^{N}\left| W_{A}-Z_{j}
\right| ^{-\frac{1}{a}\beta_{j}+\frac{\gamma_{j}}{2}}\,
\int \prod_{\alpha =1}^{s}\, \prod_{j=1}^{N}\left| U_{\alpha}-Z_{j}
\right|^{-\frac{\gamma_{j}}{2}} \nonumber \\
& & \prod_{A=1}^{t}\prod_{\alpha =1}^{s}\left| W_{A}-U_{\alpha}\right|
^{-\frac{1}{a}}\, \prod_{i\not= j}^{N}
\left| Z_{ij}\right| ^{k_{i}k_{j}+\frac{\gamma_{i}\gamma_{j}a}{4}
-\frac{\gamma_{i}\beta_{j}}{2}}.
\end{eqnarray}
We introduced the following index conventions,
\begin{equation} \label{sindexkonv}
j=1,\ldots ,N;\;  A=1,\ldots ,t;\; \alpha =1,\ldots ,s.
\end{equation}
The right hand side of (\ref{superresult}) is UV finite if
\begin{equation}     \label{sfinite}
0=k_{j}^{2}+\beta_{j}-1-\frac{\gamma_{j}\beta_{j}}{2}
+\frac{a\gamma_{j}^{2}}{4},
\end{equation}
i.e.\
\begin{equation}                                 \label{total}
\frac{1}{2}=\Delta_{j}^{0}+\Delta_{j}^{(grav)},
\end{equation}
where
\begin{equation}
\Delta_{j}^{(0)}=\frac{k_{j}^{2}}{2}
\end{equation}
is the dimension of the vertex with respect to the string part only and
(cf.\ (\ref{skondim}))
\begin{equation}
\Delta_{j}^{(grav)}=\frac{\beta_{j}}{2}-\frac{\beta_{j}\gamma_{j}}{4}+
\frac{a}{8}\gamma_{j}^{2}
\end{equation}
is the dimension of the dressing factor $B_{j}$.
The gravitational dressed dimensions are defined via the scaling behavior of
the
N--point function
\begin{equation}                 \label{suskadef}
\langle \prod_{j=1}^{N}T_{j}(Z_{j})\rangle \sim
\prod_{j=1}^{N}H^{2\Delta_{j}-1}.
\end{equation}
Hence
\begin{equation}
\Delta_{j}=\frac{1}{2}-\frac{\beta_{j}}{2}.
\end{equation}
A reasonable restriction to the special case that the dressing factor
contains only the vielbein, i.e.\
\begin{equation}
\gamma_{j}=0,
\end{equation}
leads to a trivial KPZ relation
\begin{equation}
\Delta_{j}=\Delta_{j}^{(0)}.
\end{equation}

\section{$d+2$--dimensional string}
\setcounter{equation}{0}

Let us consider the integrated N--point function,
\begin{equation}
A_{N}(k_{1},\ldots k_{N})=\frac{1}{Vol(SL(2|1))}\int
\prod_{j=1}^{N}d^{2}Z_{j}\,
\langle \prod_{j=1}^{N}T_{j}(Z_{j})\rangle .
\end{equation}
The group $SL(2|1)$, and a way to divide out it's volume are for example given
in \cite{alf}. Using the result of the previous section and neglecting
uninteresting factors provides
\begin{eqnarray}      \label{supershap}
\lefteqn{A_{N}(k_{1},\ldots ,k_{N})=}\nonumber \\
& & \frac{1}{Vol(SL(2|1))}\int \prod_{j=1}^{N}d^{2}Z_{j}\prod_{A=1}^{t}
d^{2}W_{A}\prod_{\alpha =1}^{s}d^{2}U_{\alpha} \prod_{i<j}\left|
Z_{ij}\right|^{2K_{i}K_{j}} \nonumber \\
& & \prod_{j,\alpha }\left| Z_{j}-U_{\alpha}\right|^{2iK_{j}K}
\prod_{j,A}\left| Z_{j}-W_{A}\right|^{2iK_{j}\bar{K}}
\prod_{\alpha ,A}\left| U_{\alpha}-W_{A}\right| ^{-2K\bar{K}},
\end{eqnarray}
where a capital $K$ denotes a $d+2$ dimensional vector,
\begin{eqnarray}
K_{j} & = & \left( k_{j_{1}},\ldots ,k_{j_{d}},\frac{i\beta_{j}}{2\sqrt{a}},
\frac{\sqrt{a}\gamma_{j}}{2}-\frac{\beta_{j}}{2\sqrt{a}}\right),
\\
iK & = & \left( \underbrace{0,\ldots ,0}_{d},\frac{i}{2\sqrt{a}},
-\frac{1}{2\sqrt{a}}\right),        \label{sdef-K}
\\
i\bar{K} & = & \left( \underbrace{0,\ldots ,0}_{d},\frac{i}{2\sqrt{a}},
\frac{1}{2\sqrt{a}}\right) .                  \label{sdef-barK}
\end{eqnarray}
{}From equation (\ref{sfinite}) we get
\begin{equation}
K^{2}+nK=1,
\end{equation}
with
\begin{equation}
n=\left( \underbrace{0,\ldots ,0}_{d},-2i\sqrt{a},0\right).
\end{equation}
As we will illustrate now there are two possibilities to define a mass
\begin{equation}       \label{supermasse}
-m^{2}\equiv K^{2}+nK=1,
\end{equation}
or
\begin{equation}        \label{supersnake}
-\tilde{m}^{2}\equiv \left( K+\frac{n}{2}\right)^{2}=1-a=\frac{d}{8}.
\end{equation}
Using the second definition one gets a mass less tachyon in $0+2$
dimensions ($d=0$). Let us briefly discuss equations
(\ref{supermasse}), and (\ref{supersnake}). Introducing two new
target space coordinates
\begin{eqnarray}
X^{d+1} & = & 2\sqrt{a}\left( \Sigma +\frac{i}{a}\Phi \right),\\
X^{d+2} & = & \frac{2}{\sqrt{a}}\Phi,
\end{eqnarray}
one can write the condition that the vertex has total conformal dimension
one half (\ref{total}) as follows,
\begin{equation}                                \label{supertaeq}
\left(\delta^{\mu \nu }\left(\partial_{\mu}\partial_{\nu}
+in_{\mu}\partial_{\nu}\right)
-m^{2}\right)
e^{iK_{\mu}X^{\mu}} =0.
\end{equation}
Equation (\ref{supertaeq}) is the equation of motion for the tachyon.
On the other hand one may take the point of view that the wave function
is the tachyon vertex divided by the string coupling constant
$g_{0}e^{2a\Psi}$ \cite{callan},
\begin{equation}
\tilde{T} = g_{0}^{-1}e^{-2a\Psi}e^{iK_{\mu}X^{\mu}}.
\end{equation}
In terms of $\tilde{T}$ (\ref{supertaeq}) becomes
\begin{equation}
\left( \delta^{\mu \nu}\partial_{\mu}\partial_{\nu}-\tilde{m}^{2}\right)
\tilde{T} =0.
\end{equation}
In the following we will use the mass definition (\ref{supermasse}),
a modification due to (\ref{supersnake}) is simple.
We note that one can get a Minkowskian target space instead of the Euclidean
one by putting an $i$ in front of the rhs of (\ref{sjt}).

The amplitude (\ref{supershap}) has the form of a $N+t+s$--point function
in critical string theory. An important point is that $K$, and $\bar{K}$
are zero vectors,
\begin{equation}
K^{2}=\bar{K}^{2}=0.
\end{equation}
The integrals over anticommuting coordinates $\omega_{A}$
($W_{A}=(w_{A},\omega_{A})$) do not vanish only if
\begin{equation}  \label{tleq}
t\leq s+N
\end{equation}
and integrations over $\upsilon_{\alpha}$
($U_{\alpha}=(u_{\alpha},\upsilon_{\alpha})$) yield a non zero result
only if
\begin{equation}              \label{sleq}
s\leq t+N.
\end{equation}
Hence we arrive at the condition
\begin{equation}           \label{t=s}
t=s.
\end{equation}
Since super space distances contain always pairs of anticommuting coordinates
the total number of integrations must be even. Hence a scattering amplitude of
an odd number of tachyons is always zero, i.e.\ G--parity is conserved.
Thus a GSO--projection \cite{gso} is possible although there are
background tachyons, because the background contributes with an even number
of tachyons, i.e.\ with a state of even G--parity.

Analyzing the pole structure of (\ref{supershap}) in the way described in
\cite{shap} one gets the following poles in two particle channels,
\begin{eqnarray}
S_{kl} & \equiv & \left( K_{k}+K_{l}\right)^{2}+n\left( K_{k}+K_{l}\right) =
-2j,    \label{kl-schan}           \\
S_{k\alpha} & \equiv &\left( K_{k}+iK\right)^{2}+n\left( K_{k}+iK\right)=
-2j,     \label{kalph-schan}             \\
S_{kA} & \equiv &\left( K_{k}+i\bar{K}\right)^{2}+n\left( K_{k}+
i\bar{K}\right)=-2j,   \label{kA-schan}                 \\
S_{A\alpha} & \equiv & \left( iK+i\bar{K}\right)^{2}+n\left( iK+
i\bar{K}\right)=-2j,                  \label{Aalph-schan}
\end{eqnarray}
where $j$ is a non negative integer. Equation (\ref{kl-schan}) provides
the mass spectrum of states of even G--parity, i.e.\ the spectrum which is
expected after a GSO projection is performed. Inserting for $K$, and $\bar{K}$
(\ref{sdef-K}), and (\ref{sdef-barK}) yields
\begin{eqnarray}
S_{k\alpha}& = &2-\frac{\gamma_{j}}{2}, \\
S_{kA} & = &2+\frac{\gamma_{j}}{2}-\frac{\beta_{j}}{a}, \\
S_{A\alpha} & = & -\frac{1}{a}+2.
\end{eqnarray}
Thus (\ref{kalph-schan}), and (\ref{kA-schan}) are leg poles arising due to
scattering with background tachyons of fixed momenta. Equation
(\ref{Aalph-schan}) provides poles in target space dimensions,
\begin{equation}     \label{superdimpol}
d=8\left( \frac{1+2j}{2+2j}\right), \; j=0,1,2,\ldots \; .
\end{equation}
These divergencies can be regularized by a cutoff
$|u_{\alpha}-w_{A}|>\lambda$. If we had got a result valid also for non
integer $s$, and $t$ we could perform a regularization in the target space
dimension.

Unfortunately our interpretation is strongly based on $s$, and $t$ being
integers and breaks down as soon as $s$, $t$ become real non integer
numbers.

\section{Four dimensional non critical string}
\setcounter{equation}{0}
Now we return to the non critical string, i.e.\ $\Phi$, and $\Sigma$ are not
considered as additional target space coordinates. Then it is reasonable to
restrict to the special case that gravitational dressing factors contain
only the vielbein,
\begin{equation}
\gamma_{j}=0.
\end{equation}
Moreover we are interested in a four dimensional target space,
\begin{equation}
d=4.
\end{equation}
Hence $t=1$, i.e.\ $t$ is really an integer. Instead of tachyon vertices we
prefer to take the lightest state of even G--parity,
\begin{equation}
V_{j}=\zeta^{(j)}_{\mu \nu }D_{+}X^{\mu }(Z_{j})D_{-}X^{\nu}(Z_{j})
e^{ik_{j}X(Z_{j})}B_{j}(Z_{j}),
\end{equation}
where
\begin{equation}
B_{j}(Z_{j})=\left( \epsilon e^{\Sigma (Z_{j})}\right)^{\beta_{j}-1}
e^{\Sigma (Z_{j})}.
\end{equation}
$V_{j}$ should be a primary field, i.e.\
\begin{equation}
k^{\mu}\zeta_{\mu \nu}=\delta^{\mu \nu}\zeta_{\mu \nu}=0.
\end{equation}
Analogous to the calculation described in the second section
one finds
\begin{equation}     \label{superbeta}
\beta_{j}=-k_{j}^{2}
\end{equation}
instead of
$$ \beta_{j}=1-k_{j}^{2}. $$
Using the method given in \cite{alf} to divide by the volume of $SL(2|1)$
we get for the integrated four point function
\begin{eqnarray}
\lefteqn{A_{4}=}\nonumber \\
& = & \lim_{R\to \infty}R^{2k_{4}^{2}+2+2\beta_{4}}\int d^{2}z
d^{2}\eta d^{2}\theta \, \langle V_{1}(0,0)V_{2}(z,\theta )
V_{3}(1,0)V_{4}(R,R\eta )\rangle = \nonumber \\
& = & \int d^{2}zd^{2}\eta d^{2}\theta \, \eta \theta \bar{\eta}
\bar{\theta}|z|^{2k_{1}k_{2}}|1-z|^{2k_{3}k_{2}}
\left( f_{1}\frac{1}{|z|^{2}}+f_{2}\frac{1}{|1-z|^{2}}+
\cdots \, \right) \nonumber \\
& & \int d^{2}wd^{2}\omega \, |w|^{-2\beta_{1}}
|w-z-\omega \theta |^{-2\beta_{2}}|w-1|^{-2\beta_{3}} \nonumber \\
& & |1-\eta \omega |^{-2\beta_{4}}\int \prod_{\alpha =1}^{s}
d^{2}u_{\alpha}d^{2}\upsilon_{\alpha}|u_{\alpha}-w-\upsilon_{\alpha}
\omega |^{-2}.         \label{twostar}
\end{eqnarray}
In (\ref{twostar}) $f_{1}$, and $f_{2}$ are functions of $k^{(i)}$, and
$\zeta^{(i)}$ containing no poles in the momenta.
The dots stand for a lot of additional terms which do not provide any
further information about the pole structure of $A_{4}$ \cite{doc}.

The Grassmann integrals in (\ref{twostar}) yield a non vanishing result
if
\begin{equation}
s=1.
\end{equation}
As one can convince himself by a straightforward but lengthy calculation
the same is true for terms denoted by dots. Performing integrations
over Grassmann variables we get
\begin{eqnarray}
\lefteqn{A_{4}=}\nonumber \\
& & \int d^{2}z \, |z|^{2k_{1}k_{2}}
|1-z|^{2k_{3}k_{2}}\left( f_{1}\frac{1}{|z|^{2}}+f_{2}\frac{1}{|1-z|^{2}}
+\ldots \, \right)\nonumber \\
& & \int d^{2}w\, |w|^{-2\beta_{1}}|w-z|^{-2\beta_{2}}|1-w|^{-2\beta_{3}}
\int d^{2}u \, |u-w|^{-4}.
\end{eqnarray}
The calculation of the $z$--integral is given in \cite{dotsy}. There the
two dimensional integrals are expressed by a product of contour integrals
representing the hypergeometric function. Using some identities of the
$\Gamma$--function we obtain
\begin{eqnarray}
\lefteqn{A_{4}=} \nonumber \\
& & \pi \int d^{2}w \left\{ f_{1}\left[
\frac{\Delta \left(\frac{1}{2}(k_{1}+k_{2})^{2}-\frac{1}{2}(k_{1}^{2}+
k_{2}^{2})\right) \Delta \left( 1+k_{2}^{2}\right) }{\Delta \left(
\frac{1}{2} (k_{1}+k_{2})^{2}-\frac{1}{2}(k_{1}^{2}-k_{2}^{2})\right)}\right.
\right. \nonumber \\
& & |F\left( -k_{2}k_{3}, k_{1}k_{2},1+k_{1}k_{2}+k_{2}^{2};w\right)|^{2}
|w|^{(k_{1}+k_{2})^{2}-(k_{1}^{2}-k_{2}^{2})}+  \nonumber \\
& &  +\frac{\Delta \left( \frac{1}{2} (k_{2}+k_{4})^{2}-
\frac{1}{2}(k_{2}^{2}+k_{4}^{2})\right) \Delta \left( \frac{1}{2}
(k_{2}+k_{3})^{2}-\frac{1}{2}(k_{2}^{2}+k_{3}^{2})+1\right)}{\Delta \left(
1+\frac{1}{2}(k_{2}+k_{4})^{2}+\frac{1}{2}(k_{2}+k_{3}))^{2}-\frac{1}{2}(
2k_{2}^{2}+k_{4}^{2}+k_{3}^{2})\right)}\nonumber \\
& & \left.
|F\left( -k_{2}^{2}, -k_{1}k_{2}-k_{2}k_{3}-k_{2}^{2},1-k_{1}k_{2}-k_{2}^{2}
;w\right)|^{2}\right]+ \nonumber \\
& & +f_{2}\left[ \frac{\Delta \left( \frac{1}{2}(k_{1}+k_{2})^{2}
-\frac{1}{2}(k_{1}^{2}+k_{2}^{2})+1\right) \Delta \left( 1+k_{2}^{2}\right)}
{\Delta \left( \frac{1}{2}(k_{1}+k_{2})^{2}-\frac{1}{2}(k_{1}^{2}-k_{2}^{2})
+2\right) }\right. \nonumber \\
& &
|F\left( -k_{2}k_{3}+1,k_{1}k_{2}+1,2+k_{1}k_{2}+k_{2}^{2};w\right)|^{2}
|w|^{(k_{1}+k_{2})^{2}+2-(k_{1}^{2}-k_{2}^{2})}  + \nonumber \\
& & +\frac{\Delta \left( \frac{1}{2} (k_{2}+k_{4})^{2}-
\frac{1}{2}(k_{2}^{2}+k_{4}^{2})+1\right) \Delta \left( \frac{1}{2}
(k_{2}+k_{3})^{2}-\frac{1}{2}(k_{2}^{2}+k_{3}^{2})\right)}
{\Delta \left(1+\frac{1}{2}(k_{2}+k_{4})^{2}+\frac{1}{2}(k_{2}+k_{3})^{2}
-\frac{1}{2}(2k_{2}^{2}+k_{3}^{2}+k_{4}^{2})\right)}\nonumber \\
& & \left. \left.
|F\left( -k_{2}^{2},-k_{1}k_{2}-k_{2}k_{3}-k_{2}^{2},-k_{1}k_{2}-k_{2}^{2};
w\right) |^{2} \right]+\ldots \right\} \nonumber \\
& &  |w|^{2k_{1}^{2}}|1-w|^{2k_{3}^{2}}\int d^{2}u \, |u-w|^{-4}.
\end{eqnarray}
$F$ is the hypergeometric function and
\begin{equation}
\Delta (x)=\frac{\Gamma (x)}{\Gamma (1-x)}.
\end{equation}
After a shift of $u$ by $w$ the $u$--integral factorizes providing a divergent
factor. However, since our calculation is valid for $s=1$ the partition
function contains the same divergent factor. Hence the divergency cancels
due to normalization.

For fixed $k^{2}_{i}$ the following poles of two particle channels occur,
\begin{equation}                      \label{superpole1}
(k_{1}+k_{2})^{2}=-2j+k_{1}^{2}+k_{2}^{2}, \; \; j=0,1,2,\ldots,
\end{equation}
\begin{equation}                        \label{superpole2}
(k_{2}+k_{3})^{2}=-2j+k_{2}^{2}+k_{3}^{2},
\end{equation}
\begin{equation}                        \label{superpole3}
(k_{2}+k_{4})^{2}=-2j+k_{2}^{2}+k_{4}^{2}.
\end{equation}
Furthermore we observe a leg pole,
\begin{equation}       \label{superbeinpol}
k_{2}^{2}=-1-j.
\end{equation}
As one can convince himself by an expansion of the integrand in a power series
around $|w|=0$, and $|w|=\infty$ there are no further poles of the
types (\ref{superpole1})--(\ref{superpole3}).

As in the purely bosonic case \cite{me} we suppose that we are given a
mass $m_{0}$ of the ground state
from somewhere else and consider on shell amplitudes,
\begin{equation}
k_{j}^{2}=-m_{0}^{2}.
\end{equation}
Then we get poles e.g.\ for
\begin{equation}                   \label{doppelkreis}
(k_{1}+k_{2})^{2}=-2j-2m_{0}^{2}.
\end{equation}
For consistency we require the amplitude to possess poles at
$$ (k_{1}+k_{2})^{2}=-m_{0}^{2}.  $$
(We note that this requirement is reasonable because we consider vertices
of even G--parity.)
That leads to a restriction of $m_{0}$,
\begin{equation}        \label{kreuz}
m_{0}^{2}=-2j_{0},
\end{equation}
where $j_{0}$ is an arbitrary fixed non negative integer.
Thus we get no additional divergencies due to leg poles (\ref{superbeinpol}).
With (\ref{doppelkreis}) we get the full mass spectrum,
\begin{equation}
M_{j}^{2}=2j-4j_{0}.
\end{equation}
Now we require the ground state to be the lightest one,
\begin{equation}
m_{0}^{2}\leq M_{j}^{2},\; \; \forall j.
\end{equation}
That leads to $j_{0}=0$, i.e.
\begin{equation}                   \label{superspeck}
M_{j}^{2}=2j,\; \; j=0,1,2,\ldots .
\end{equation}
With (\ref{superspeck}) follows
\begin{equation}
s=1.
\end{equation}
Although we treated $s$ like an independent parameter during the calculation
our result is self consistent, at least on shell.
(It is also self consistent if the sum of squared momenta is the same
as it would be if all single momenta were on shell.)

In on shell amplitudes the gravitational expectation value factorizes because
$\beta_{j}=0$. Hence on shell amplitudes in four dimensional non critical
string theory correspond to those of ten dimensional critical string theory.
Thus a GSO--projection is possible also in four dimensions.
Off shell states are accompanied by two dimensional gravitons, and
gravitinos which interact with background gravitons, gravitinos, and
dilatons. The background particles occur because the curvature of the
world sheet is not zero and hence there must be a force creating a
non vanishing curvature. (In our model zero curvature would imply $H=0$
and, in fact, for $H=0$ there were no background particles.)
Since momenta of background particles are fixed we get leg poles
(\ref{superbeinpol}).

\section{Conclusions}
\setcounter{equation}{0}

Taking into account the Jackiw--Teitelboim action as an action for pure
2D super gravity we obtained a trivial KPZ relation. Hence scaling
dimensions of vertex operators are always real. The Jackiw--Teitelboim
action creates the constraint of constant curvature and trivializes in
that way Liouville quantization. However, in order to calculate correlation
functions one has to use translation invariant measures, i.e.\ to split the
vielbein into a quantum part and a background part. Then the constraint of
constant curvature is no longer valid for the quantum field and the
trivialization disappears in intermediate calculations. But the lack of
renormalization of conformal dimensions is a hint that our calculation
is correct, nevertheless.

In order to get further results one has to restrict to the special case
that $s$, and $t$ are non negative integers. Unfortunately up to now we
have not been able to express arising two dimensional integrals in such a way
that a continuation to real values of $s$, and $t$ would be possible. The
method described in \cite{dotsi} is applicable only for certain parameters.
In our case we would have to chose $d=4$. But for $d=4$ our result is
divergent (cf.\ (\ref{superdimpol})) and has to be regularized in an
appropriate way.

In the $d+2$--dimensional critical string picture another interesting special
case would be $d=0$. Then our model described the two dimensional critical
string and a comparison with results obtained in a calculation without
Jackiw--Teitelboim action \cite{apiky} would be possible. For $d=0$ there
were no string coordinates and hence no Liouville action would be induced.
Therefore we would have to calculate with $d>0$ and perform the limit
$d\rightarrow 0$ afterwards. Since our result makes sense only for integer
$s,t$ it seems not to be reasonable to perform that limit.

However, in the region where our calculation is valid it provides interesting
statements. Although there are background tachyons a GSO projection is
possible because the background takes part in scattering always with an
even number of tachyons and hence does not violate G--parity conservation.

Interpreting $\Phi$, and $\Sigma$ not as additional target space coordinates
we obtained a mass spectrum via consistency requirements. It is the same
as in ten dimensional critical string theory. Moreover on shell amplitudes
are equal to critical string amplitudes. Hence in that picture a GSO
projection is possible, too.

It would be interesting to take into account the Ramond sector. Then we
were able to describe half integer spin excitations, too \cite{gso}.
If then to every bosonic degree of freedom belonged a fermionic one
that would be a first hint that a construction of a non critical four
dimensional space time super string should be possible.\\ \mbox{}\\
\begin{center}{\bf Acknowledgements}\end{center}
I thank H.-J. Otto for fruitful discussions.
I obtained useful hints also in discussions with H.\ Dorn, Vl.\ S.\ Dotsenko
and S.\ D.\ Odintsov.\\

\end{document}